\documentclass[twoside,a4paper,11pt]{torus2012}
\usepackage{graphicx}
\usepackage{hyperref}
\usepackage{movie15}

\def\apj{\mbox{ApJ}}
\def\apjl{\mbox{ApJL}}
\def\apjs{\mbox{ApJS}}

\def\mnras{\mbox{MNRAS}}

\def\aap{\mbox{A\&A}}
\begin{document}
\pagenumbering{arabic}
\pagestyle{myheadings}
\thispagestyle{empty}
{\flushleft\includegraphics[width=\textwidth,bb=58 650 590 680]{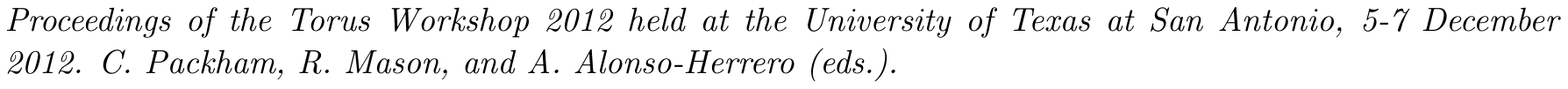}}
\vspace*{0.2cm}
\begin{flushleft}
{\bf {\LARGE
%
%%% TITLE of the paper. 
%%% TITLE of the paper. 
The AGN dusty torus as a clumpy two-phase medium: radiative transfer
modeling with SKIRT
%
% Do not delete next few lines
}\\
\vspace*{1cm}
%
%%% Include here the LIST OF AUTHORS.
%%% Include here the LIST OF AUTHORS.
%%% Note that the last author has to be preceeded by an AND.
Marko Stalevski$^{1,2}$,
Jacopo Fritz$^{2}$, 
Maarten Baes$^{2}$,
and 
Luka \v C. Popovi\'{c}$^{1,3}$
%
% Do not delete next few lines
}\\
\vspace*{0.5cm}
%
%%% AFFILIATIONS LIST.
%%% and the AFFILIATIONS LIST. Note that one affiliation per line.
%%% Add as many affiliations as necessary. 
$^{1}$
Astronomical Observatory, Volgina 7, 11060 Belgrade, Serbia\\
$^{2}$
Sterrenkundig Observatorium, Universiteit Gent, Krijgslaan 281-S9,
Gent, 9000, Belgium\\
$^{3}$
Department of Astronomy, Faculty of Mathematics, University of
Belgrade
%
% Do not delete next few lines
\end{flushleft}
%
% Headings
\markboth{
%%% Type the SHORT version of the paper title.
%%% Type the SHORT version of the paper title.
The AGN dusty torus as a clumpy two-phase medium
}{ % Do not delete
%
%%%  First Author \& Second Author   OR   First-author et al. 
%%%  First Author \& Second Author   OR   First-author et al. if the author list 
%%% contains three or more authors.
Marko Stalevski et al.
% 
% Do not delete next few lines
}
\thispagestyle{empty}
\vspace*{0.4cm}
\begin{minipage}[l]{0.09\textwidth}
\ 
\end{minipage}
\begin{minipage}[r]{0.9\textwidth}
\vspace{1cm}
\section*{Abstract}{\small
%
% ABSTRACT ABSTRACT ABSTRACT
% ABSTRACT ABSTRACT ABSTRACT
%%% Type the ABSTRACT of your paper
We modeled the AGN dusty torus as a clumpy two-phase medium, with high-density
clumps embedded in a low-density interclump dust. To obtain spectral
energy distributions and images of the torus at different wavelengths,
we employed the 3D Monte Carlo radiative transfer code \textsc{skirt}.
Apart from the grid of two-phase models, we calculated the
corresponding sets of clumps-only models and models with a smooth
dust distribution for comparison. We found that the most striking
feature of the two-phase model is that it might offer a natural
solution to the common issue reported in a number of papers --- the
observed excess of the near-infrared emission.
%
% Do not delete next few lines
\normalsize}
\end{minipage}
%
%
%%% BODY of the paper
%%% BODY of the paper
%
%--------------------------------------------------------------------
\section{Introduction }
\label{intro}
%--------------------------------------------------------------------
The unification model of active galactic nuclei (AGN) requires a
roughly
toroidal structure
of dust and gas that surrounds the central regions. In order to prevent the dust
grains from being destroyed by the hot surrounding gas, it has been suggested
that the dust in the torus is organized in a large number of optically thick
clumps \cite{krolikbegel88}. However, due to the difficulties in handling clumpy
media and lack of computational power, early works adopted a smooth dust
distribution, e.g. \cite{pierkrolik92, granatodanese94, fritz06}. Later on,
several authors undertook different efforts for the treatment of clumpy media,
e.g. \cite{honig06, nenkova08a, schartmann08} (for an overview of
different AGN torus models see contribution by
Sebastian H\"{o}nig in this proceedings, \cite{Hoenig13}). A problem
which the obscuring torus
hypothesis faced from the beginning was formation of the dynamically stable
structure and maintenance of the required scale-height. According to one
scenario, the scale-height is maintained through regular elastic collisions
between the clumps \cite{krolikbegel88}. In the case of a smooth dust
distribution, radiation pressure within the torus may be enough to
support the
structure \cite{pierkrolik92, krolik07}. Hydrodynamical simulations, taking into
account processes as self-gravity of the gas, radiative cooling and
heating due to supernovae, found that such a turbulent
interstellar medium in AGN would result in a multiphase filamentary
(sponge-like) structure, rather then isolated clumps
\cite{wada09,Wada12}. Although the current observational facilities
can resolve the structure of only a few nearby AGN tori, there are
indications that it indeed consists of thick clumps embedded in a
diffuse interclump medium \cite{Assef12}.
Also, such two-phase structures have been actually observed in the
central
regions of
Milky Way (the so-called Central Molecular Zone and Circumnuclear Disk) and it
has been suggested that they represent a remnant of a dusty torus that may have
played a role in past AGN phases of our Galaxy \cite{Molinari11,Ponti12}. Our
aim was to take a step further toward a more realistic model by treating the
dusty torus as a two-phase medium, with high density clumps and low density
medium filling the space between them. To calculate spectral energy
distributions (SEDs) and images of the
torus, we used the 3D Monte Carlo radiative transfer code \textsc{skirt}
\cite{baes03,baes11}. The approach we adopted for building our model enables us
to, for each two-phase model, calculate a corresponding clumps-only model (with
dust distributed to the clumps exclusively, without any dust between them) and a
smooth model, with the same global physical parameters, allowing a consistent
comparison between them \cite{stalevski12a}. 
%--------------------------------------------------------------------
\section{Model}
%--------------------------------------------------------------------
\subsection{Dust distribution and properties}

For the detailed description of the model and its parameters, we refer the
reader to \cite{stalevski12a}; here we will present only the most important
properties and general approach.
The primary continuum source of dust heating is the intense UV-optical continuum
coming from the accretion disk. A very good approximation of its emission is a
central, point-like energy source, whose SED is very well approximated by a
composition of power laws \cite{stalevski12a}. We approximate the
spatial dust distribution around
the primary continuum source with a conical torus (i.e.~a flared disk), whose
characteristics are defined by the half opening angle and the inner and outer
radii. For the inner radius we adopted the value of $0.5$ pc, at wich the dust
grains are heated to the temperature of $\sim 1180$ K, according to the
prescription given by \cite{barvainis87}. We describe the spatial distribution
of the dust density with a law that allows a density gradient along the radial
direction and with polar angle. The dust mixture consists of separate
populations of graphite and silicate dust grains with a classical MRN size
distribution \cite{mrn77}.

The dust is distributed on a 3D cartesian grid composed of a large
number of cubic cells. The standard resolution for our simulations is
200 cells along each
axis ($8\times10^{6}$ cells in total). To ensure that the adopted
resolution is sufficient to properly sample the dust, for each
simulation we compare the actual and expected values of (a)\ the
face-on and edge-on central surface density and (b)\ the total dust
mass. The emission for all models was calculated on an equally spaced
logarithmic wavelength grid ranging from $0.001$ to $1000$ $\mu$m. A
finer wavelength sampling was adopted between $5$ and $35$ $\mu$m, in
order to better resolve the shape of $10$ and $18$ $\mu$m silicate
features. Each simulation is calculated using $10^{8}$ photon
packages.

%--------------------------------------------------------------------
\subsection{Two-phase medium: the approach}
%--------------------------------------------------------------------
%---------------------------------
\begin{figure}
\centering
\includegraphics[scale=0.22]{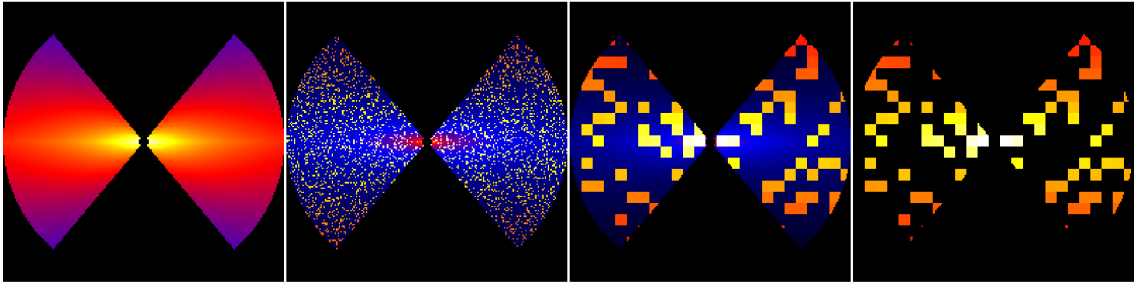}
\caption{\label{fig1} Dust
density distribution in the meridional plane, given in
logarithmic color scale. The smooth dust distribution is presented in the
leftmost panel. The two middle panels present two-phase dust
distribution for two different sizes of clumps: each clump is
composed of one cubic grid cell and $8\times8\times8$
grid cells . The rightmost panel shows a clumps-only
dust distribution. The contrast parameter between high-
and low-density phases in the two-phase and clumps-only models is
$100$ and $10^9$, respectively.}
\end{figure}
%---------------------------------
In the approach we adopted to build a two-phase model, the dust is
first
distributed smoothly within the
toroidal region, according to the adopted density law. The total amount of dust
is fixed based on the equatorial optical depth at $9.7$ $\mu$m.
Then, to generate a two-phase clumpy medium, we apply the algorithm described by
\cite{wittgordon96}. According to this algorithm, each individual
cell in the grid is assigned randomly to either a high- or
low-density state by a Monte Carlo process. The medium created in
such a way is statistically homogeneous, but clumpy. The filling
factor determines the statistical frequency of the cells in the
high-density state and can take values between $0$ and $1$. For
example, a filling factor of $0.01$ represents a case of rare, single
high-density clumps in an extended low-density medium. The process
for the clump generation is random with respect to the spatial
coordinates of the individual clumps themselves. Thus, as the filling
factor is allowed to increase, the likelihood that adjoining cells
are occupied by clumps increases as well. This leads to the
appearance of complex structures formed by several merged clumps. For
filling factor values $>0.25$, clumps start to form an
interconnected sponge-like structure, with low-density medium filling
the voids. We form larger clumps by forcing several adjoining cells
into a high-density state. To tune the density of the clumps and the
inter-clump
medium, we use the `contrast parameter', defined as the ratio of the dust
density in the high- and low-density phase. This parameter can be assigned
any positive value. For example, setting the contrast to one would
result in a smooth model. Setting extremely high value of contrast
($>1000$) effectively puts all the dust into the clumps, without
low-density medium between them. An example of dust density
distributions in the meridional plane for smooth, two-phase and
clumps-only models is given in Fig.~\ref{fig1}.

%--------------------------------------------------------------------
\section{Results and Discussion}
\begin{figure*}
\centering
\includegraphics[scale=0.3]{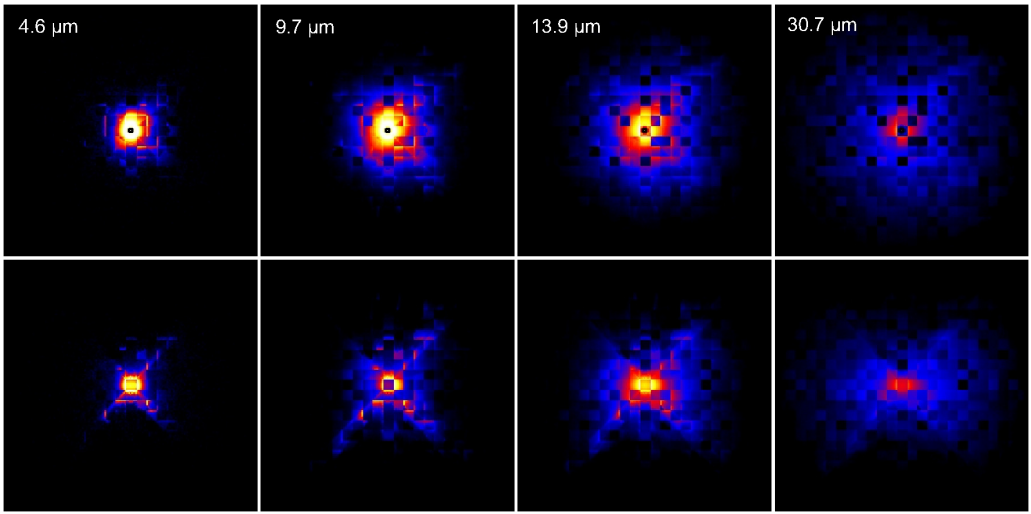}
\caption{\label{fig2} Images of the
torus at different wavelengths, in logarithmic color scale. Top row is
face-on view, bottom row edge-on view. The visible squared structure is due
to clumps which in the model are in the form of cubes.}
\end{figure*}

%--------------------------------------------------------------------
An example of images of the torus for typical parameters at different
wavelengths is shown in
Fig.~\ref{fig2}. Movies, in which each frame represents an image
of the torus at different wavelength between $0.1$ and $1000$ $\mu$m,
are available at following hyper-links:
\href{http://goo.gl/aRZPL}{face-on} and
\href{http://goo.gl/8BOqb}{edge-on} view (note that the images
presented in these movies were calculated with updated model where
clumps are in the form of spheres). It is apparent that the size of
the torus is wavelength
dependent: at shorter wavelengths, it is the radiation from the inner (and
hotter) region that dominates, while at longer wavelengths, the emission arises
from the dust placed further away. A detailed analysis of the properties of the
dusty torus infrared emission can be found in \cite{stalevski12a}. In
the remainder
of this contribution, we will bring to focus only several of the most
interesting results. A grid of calculated SEDs is publicly
available at \url{https://sites.google.com/site/skirtorus/}.
%---------------------------------
%---------------------------------
%--------------------------------------------------------------------
\subsection{Two-phase medium: a natural solution to the near-IR
excess?}
%--------------------------------------------------------------------
%---------------------------------
\begin{figure*}[h]
\centering
\includegraphics[scale=0.6]{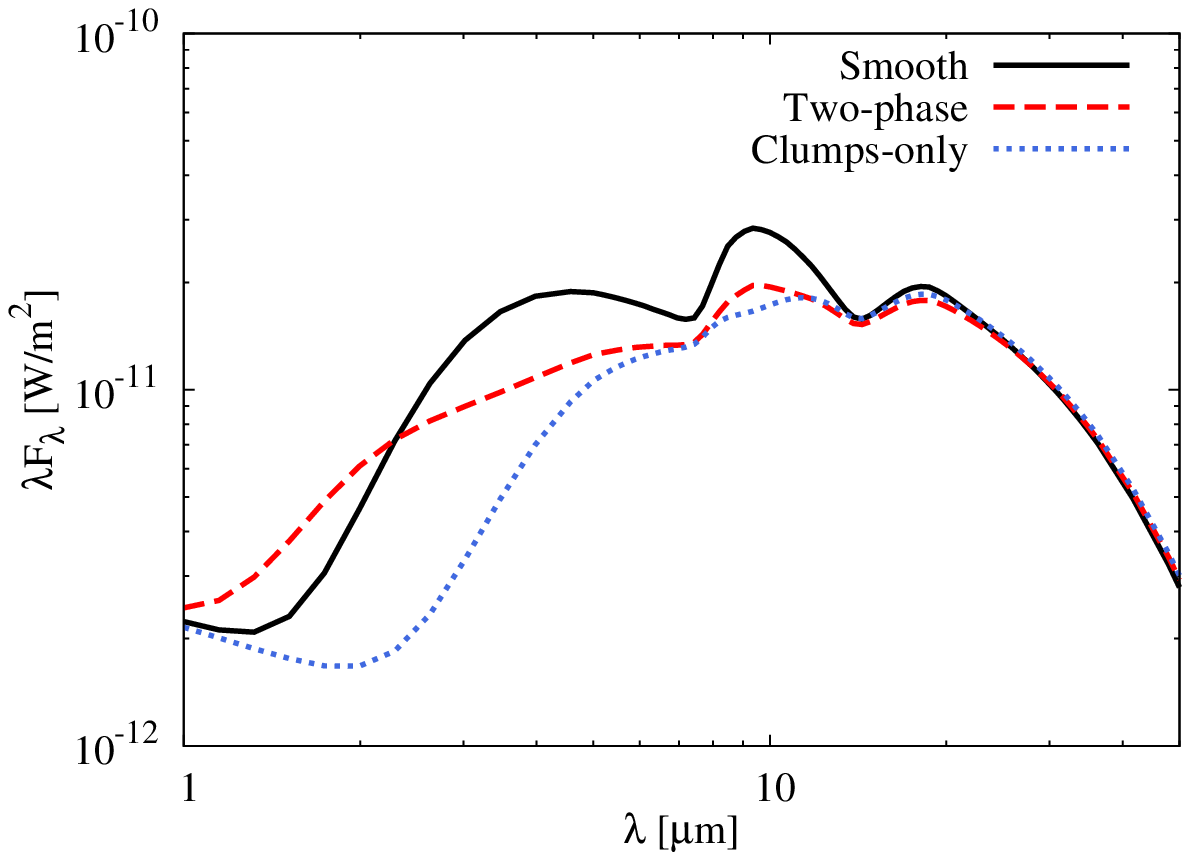}
\includegraphics[scale=0.6]{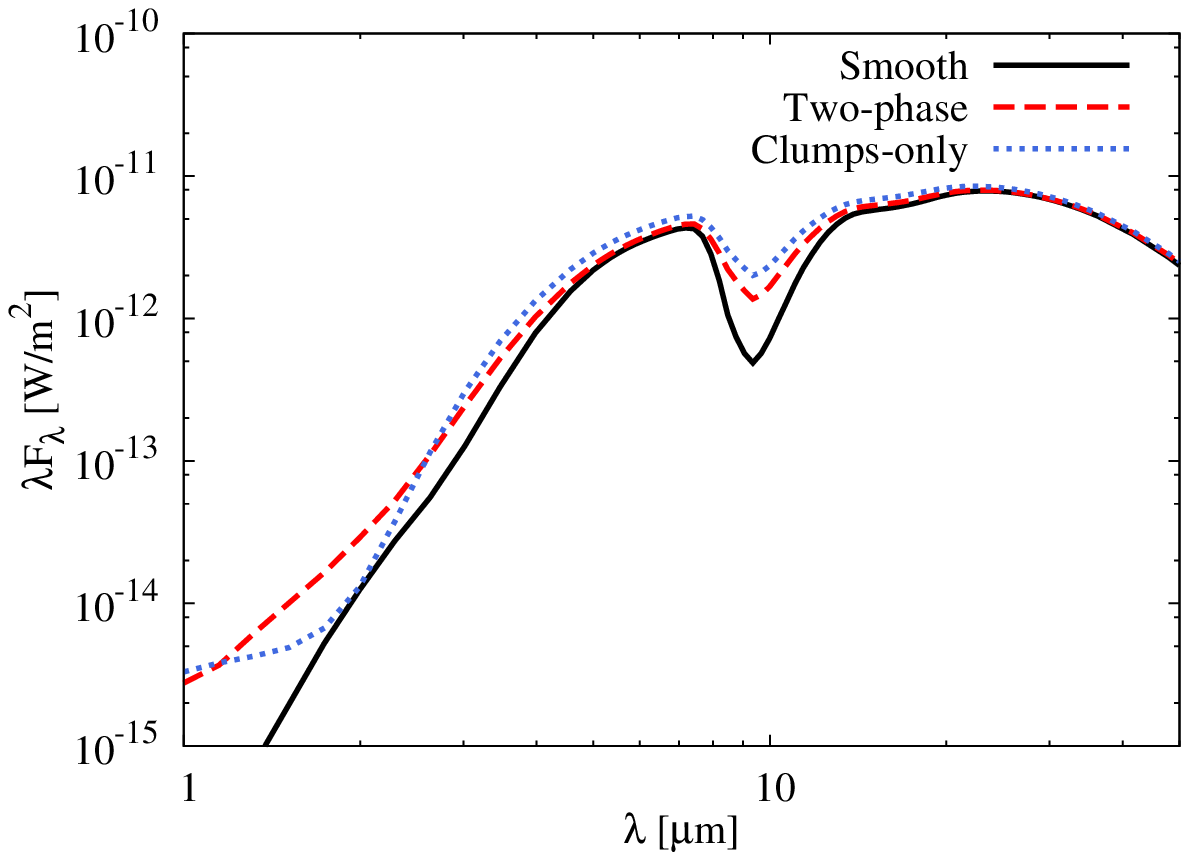}
\caption{\label{fig3} Comparison of
the smooth (solid), two-phase (dashed) and clumps-only (dotted) model
SEDs for typical torus parameters, for face-on (left panel) and
edge-on (right panel) view.}
\end{figure*}
%---------------------------------
In Fig.~\ref{fig3}, SEDs of smooth, two-phase and clumps-only models
are
compared. The major difference between SEDs of two-phase and clumps-only models
arises in the near-IR range and mainly for the face-on view. At these
wavelengths, the most of the two-phase models with the type 1 
inclination have a flatter SED when compared to the corresponding clumps-only
models. This difference is caused by the presence of the smooth
component in which the clumps are embedded. Dust in this component can reach
high temperatures and will give rise to the higher luminosity in the $2-6\,
\mu$m range. Regarding the $10$ $\mu$m silicate feature, we do not find any
significant difference between the two dust configurations: depending
on parameters, in clumps-only models it could be slightly
attenuated compared to the one in the two-phase models, but the
difference is in most cases marginal. A similar behavior can be
observed in the SEDs corresponding to edge-on views, in which the
smooth low-density component is responsible for an additional
absorption, so the silicate feature is slightly deeper in the
two-phase models.

It is very interesting to note that such a behavior of the
near- and mid-IR emission of the two-phase dust distribution, may
in some cases overcome an issue that seems to be common to the clumpy
models
currently available in the literature. A number of papers studying properties
of infrared emission of different samples of AGNs reported the need of
an extra hot-dust component to reproduce observed SEDs
\cite{polletta08, mor09, deo11, vignali11}. In these cases, adopted
clumpy models were not able to simultaneously reproduce the intensity
of the silicate feature and the near-IR continuum emission: models
that would properly fit the near-IR were overestimating the silicate
feature emission. It has been suggested that graphite at the
sublimation limit could be responsible for the extra near-IR emission.
Although this graphite is a plausible source of an additional near-IR
emission, it has yet to be confirmed by modeling. Rather than simply
adding a black-body component to the fitting procedure, the dust
component responsible for the observed near-IR excess must be
consistently included and present in the model during the radiative
transfer simulation itself. 

On the other hand, as we have shown, a torus model with the dust distributed in
a two-phase medium, has a more pronounced (`hotter') emission in the
$2-6$ $\mu$m range while displaying, at the same time, a silicate
feature whose intensity is almost identical to that of the
corresponding clumps-only model. Thus, it seems that the two-phase models may
offer a natural solution to the observed excess of near-IR emission. This is
supported by a recent study by \cite{roseboom12}. The authors considered the
optical to
mid-IR properties of a sample of quasars and estimated a number of
properties, including the IR SED shape characterized by the ratio of
near-IR ($1-5$ $\mu$m ) to total IR luminosity. The typical ratio
of the near-IR to the total IR luminosity in their sample is 
$\sim40$~\%. They found that this ratio is achievable in the CLUMPY
models \cite{nenkova08a} only in a limited range of parameter values.
On the other hand, in the set of two-phase models, they found that
near-IR to total IR luminosity ratios similar to that in the observed
sample are easily achievable.
%--------------------------------------------------------------------
\subsection{Anisotropic accretion disk emission}
\begin{figure}[h]
\centering
\includegraphics[scale=0.7]{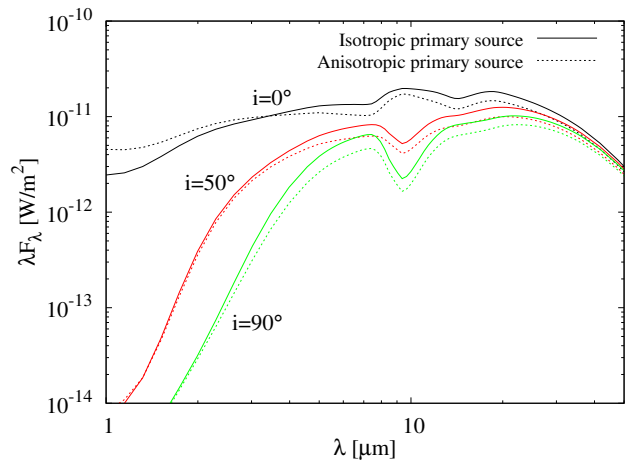}
\caption{\label{fig4} Model SEDs assuming isotropic (solid line)
and anisotropic
(dotted line) accretion disk radiation. Inclination of $i=0^\circ$ represents a
face-on view, $i=90^\circ$ an edge-on view.}
\end{figure}

%--------------------------------------------------------------------
An isotropic source emission is commonly adopted in the radiative transfer
modeling of the dusty tori. However, owing to the change in the
projected
surface area and limb darkening, the accretion disk emission is
actually anisotropic \cite{netz87}. We took this into account as a
$cosine$ dependence of
the accretion disk radiation on the direction and with the corresponding change
of
the dust sublimation radius \cite{stalevski12a}. In
Fig.~\ref{fig4} we present the resulting model SEDs if the
anisotropic radiation of the primary source is assumed (dotted line)
and compare them to the corresponding SEDs obtained in the case of
the isotropic source (solid line) for different inclinations.
We found that, when anisotropy of the central source is assumed, the
infrared SED can indeed change, resulting in a lower emission, though
roughly keeping the same shape. This is a logical consequence coming
from the fact that, for a given bolometric luminosity of the
accretion disk, an anisotropic source is emitting more power in the dust-free
region: the overall result is a less luminous torus. We found that the shape and
the features of the SED (e.g. the $10\, \mu$m feature) are only marginally
affected. It is interesting to note the additional emission shortward of
$\sim3\, \mu$m, seen in the dust-free lines of sight. However, in the current
simulations, we were limited by the memory requirements; a better resolution in
the inner regions is needed to study the effects of anisotropic source and
possible relevance for the issue of the near-IR excess. The adaptive
octree grid
that is implemented in the \textsc{skirt} code \cite{saftly13} will allow us to
do this.
%--------------------------------------------------------------------
\subsection{Degeneracy due to the random distribution of clumps}
%--------------------------------------------------------------------
%---------------------------------
\begin{figure*}
\centering
\includegraphics[height=0.23\textwidth]{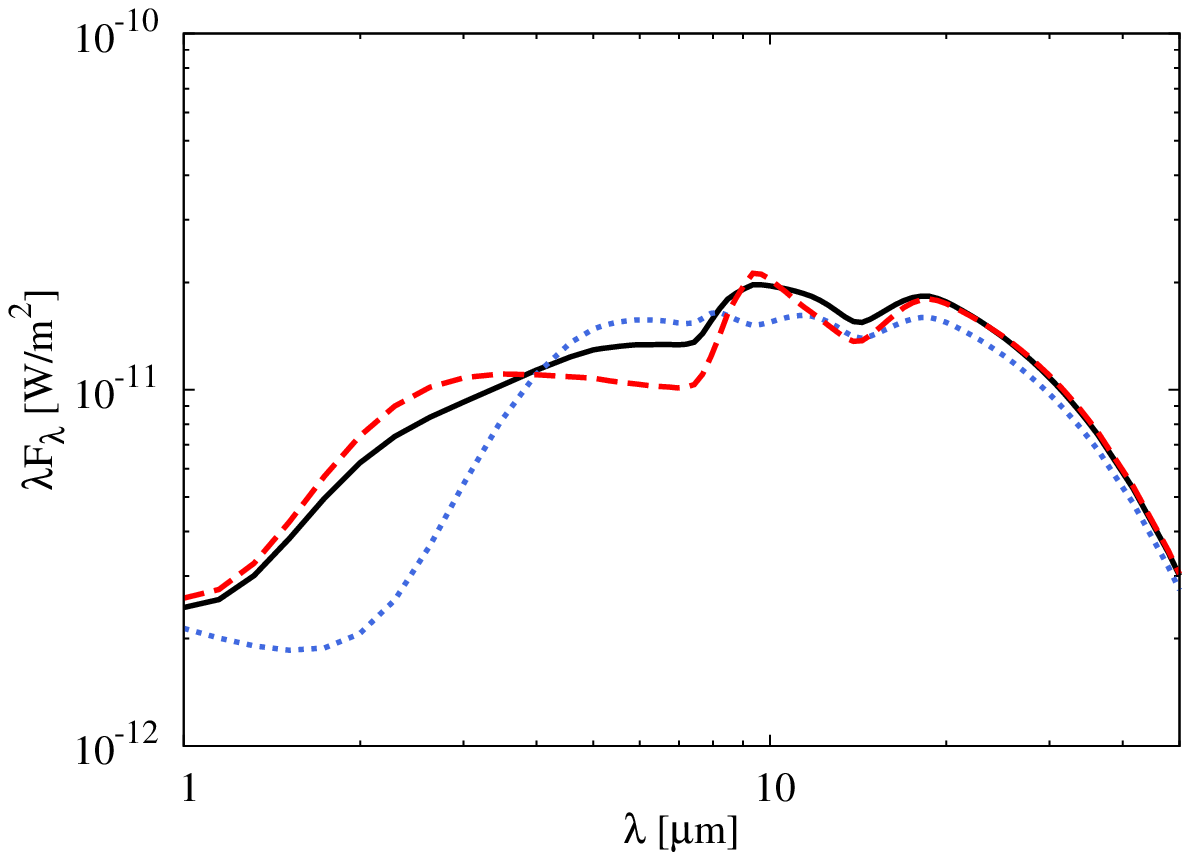}
\includegraphics[height=0.23\textwidth]{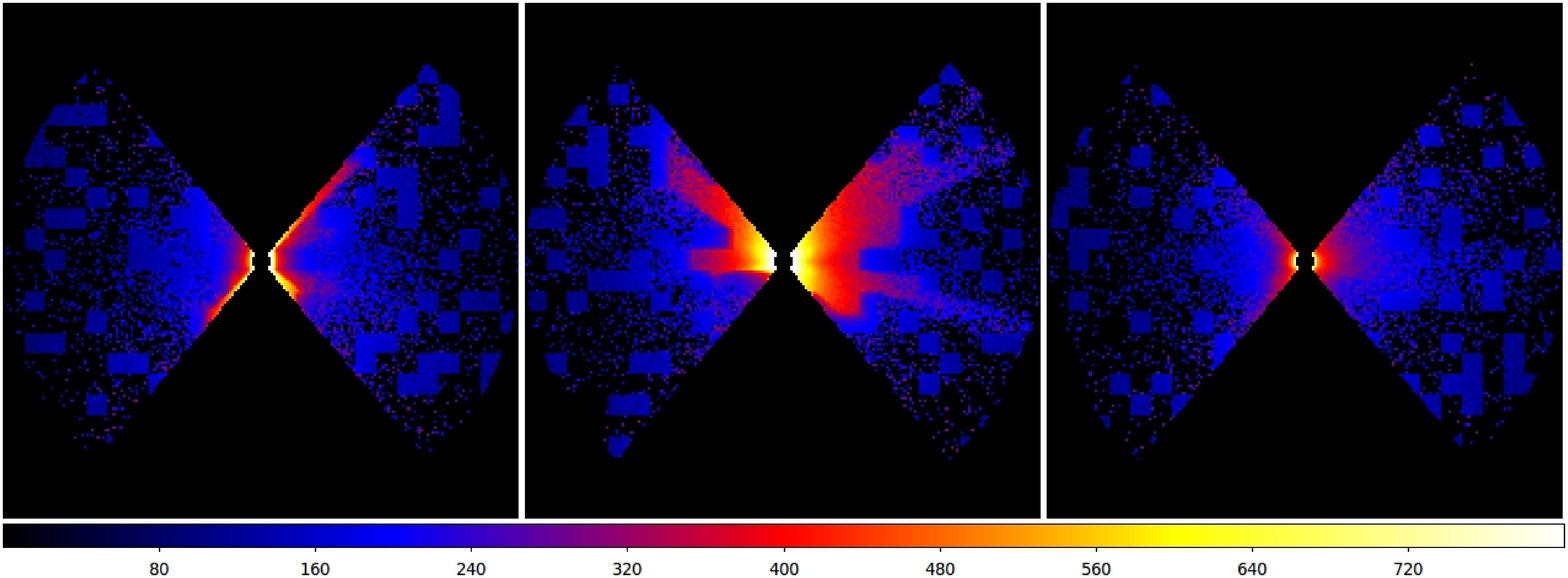}
\caption{\label{fig5} Temperature distribution in the meridional plane
for the three different random distributions of
clumps (three rightmost panels) and corresponding SEDs for the face-on
view (left panel). The solid line corresponds to the left panel, the
dashed to the middle, and the dotted to the right. Temperature
distributions are shown in a logarithmic color scale, which is for
clarity of the images cut off at $800$ K.}
\end{figure*}
%---------------------------------
The shape and overall near- and mid-IR emission strongly depend
on the distribution of dust in the innermost region. Changing the
random arrangement of clumps, along with choosing a particular line
of sight, can affect the resulting SED significantly, as illustrated
in Fig.~\ref{fig5}. This characteristic imports a degree of the
degeneracy in
the features of SEDs, which will depend less directly on the
physical input parameters. Even though the spatial position of the
clumps is not related to the physical properties of the dusty tori,
their re-arrangement has a clear impact on the infrared emission. It
is, in some way, mimicking a change in the optical depth, which might
appear either to increase or decrease, depending
on the clumps re-arrangement, especially in the innermost regions. For example,
in models with a higher concentration of clumps in the innermost region, due to
the shadowing effect, the absorption is increased and the silicate
feature is suppressed. However, we stress that the SEDs shown
in Fig.~\ref{fig5} present the extreme examples; many other random
distributions of clumps would not result in such a drastic differences. Also,
due to the tidal shearing, clumps at the inner rim should be smaller than
clumps further away. This could potentially reduce the differences seen in
SEDs, as they are most affected by the dust properties in the innermost
region. Further simulations with higher resolution will enable us to
estimate the degree and importance of degeneracy introduced by random
arrangements of clumps. 
%
%
% Do not delete the next line
\small  % Do not delete
%
%%% Comment the following line if you do not have acknowledgments.
\section*{Acknowledgments}   % Do not delete if you declare acknowledgments
%
%%% ACKNOWLEDGMENTS
%%% ACKNOWLEDGMENTS
This work was supported by the Ministry of Education, Science and
Technological Development of the Republic of Serbia through the
projects ‘Astrophysical Spectroscopy of Extragalactic Objects’
(176001) and ‘Gravitation and the Large Scale Structure of the
Universe’ (176003).

%
% Do not delete the next few lines

%
\end{document}